\documentclass[aps,pra,twocolumn,showpacs,superscriptaddress]{revtex4-1}
\usepackage[colorlinks ,linkcolor=blue,anchorcolor=blue,citecolor=blue,urlcolor=blue]{hyperref}
\usepackage{graphicx}
\usepackage{epstopdf}
\usepackage{soul, color, xcolor}
\usepackage{bm}
\usepackage{amsmath,amssymb,amsfonts}
\usepackage{appendix}

\def\be{\begin{equation}} \def\ee{\end{equation}}
\def\bea{\begin{eqnarray}} \def\eea{\end{eqnarray}}

\def\nn{\nonumber}

\begin{document}

\title{Nonlinear valley Hall effect in a bilayer transition metal dichalcogenide}
\author{Zhichao Zhou}
\affiliation{School of Physics and Technology, Nanjing Normal University, Nanjing 210023, China}
\author{Ruijing Fang}
\affiliation{School of Physics and Technology, Nanjing Normal University, Nanjing 210023, China}
\affiliation{Center for Quantum Transport and Thermal Energy Science,
Nanjing Normal University, Nanjing 210023, China} 
\author{Zhen Zhang}
\affiliation{School of Physics and Technology, Nanjing Normal University, Nanjing 210023, China}
\affiliation{Center for Quantum Transport and Thermal Energy Science,
Nanjing Normal University, Nanjing 210023, China} 
\author{Xiaoyu Wang}
\affiliation{School of Physics and Technology, Nanjing Normal University, Nanjing 210023, China}
\affiliation{Center for Quantum Transport and Thermal Energy Science,
Nanjing Normal University, Nanjing 210023, China} 
\author{Jiayan Rong}
\affiliation{School of Physics and Technology, Nanjing Normal University, Nanjing 210023, China} 
\author{Xiao Li} 
\email{lixiao@njnu.edu.cn}
\affiliation{School of Physics and Technology, Nanjing Normal University, Nanjing 210023, China}
\affiliation{Center for Quantum Transport and Thermal Energy Science,
Nanjing Normal University, Nanjing 210023, China}

\begin{abstract} 
Valley-contrasting Hall transport conventionally relies on the inversion symmetry breaking in two-dimensional systems, which greatly limits the selection range of valley materials. 
In particular, while monolayer transition metal dichalcogenides have been widely utilized as a well-known class of valley materials in valleytronics, the centrosymmetric nature hinders the realization of valley-contrasting properties in the bilayer counterparts.     
Here, taking MoS$_{2}$ as an example, we discover valley-contrasting transport in bilayer transition metal dichalcogenides by exploring nonlinear transport regime. 
 Using effective models and first-principles calculations, our work demonstrates that nonvanishing nonlinear valley Hall conductivities emerge in a uniaxially strained MoS$_{2}$ bilayer, owing to strain-induced band tilts of Dirac fermions.
 With the aid of small spin-orbit-coupling induced band splittings, the conduction bands generate much remarkable nonlinear valley Hall conductivity. 
Moreover, the nonlinear conductivities are highly tunable through modulating the strength and the direction of the strain, chemical potential, and interlayer gap. 
Our findings not only expands material choices for valleytronic applications, but also provides opportunities for designing advanced electronic devices that leverage nonlinear valley transports. 
\end{abstract} 
\maketitle

{\color{blue}\textit{Introduction.}} -- Valley, as a local extreme in band structure, is an emergent electronic degree of freedom \cite{Xiao2007, Schaibley2016}. 
Inequivalent valleys possess valley-contrasting transport and optoelectronic properties \cite{Xiao2007, Yao2008}. 
For example, under the action of in-plane electric field, opposite anomalous Hall currents driven by opposite Berry curvatures are generated from different valleys, leading to valley Hall effect. 
With the help of valley-contrasting properties, valley degree of freedom has enormous potential applications in information encoding and manipulation. 
To realize valley-contrasting properties, the broken spatial inversion symmetry is regarded as a necessity.  
Therefore, while noncentrosymmetric monolayers of transition metal dichalcogenides, MX$_{2}$ (M=Mo,W, etc.; X=S, Se, etc.), are a class of typical valley materials \cite{Xiao2012, Cao2012, Zeng2012}, valley-contrasting properties are absent in readily available centrosymmetric MX$_{2}$ bilayers and graphene bilayer, unless an external stimuli is introduced to break the symmetry \cite{Wu2013, Xiao2007, Sui2015, Shimazaki2015}.  
The symmetry constraint greatly limits material candidates that exhibit valley-contrasting properties.

The above symmetry requirement is established based on the linear valley response to external perturbations. 
Given that nonlinear transport effects have recently been arising topics in condensed matter physics \cite{Ideue2021, Du2021, Bandyopadhyay2024}, they provide opportunities for boosting valley physics beyond the linear response regime. 
Different from the linear counterpart, the leading contribution of the nonlinear Hall transport to transverse currents  quadratically depends on the longitudinal electric field, and arises from the Berry curvature dipole \cite{Sodemann2015, Du2018} or Berry connection polarizability \cite{GaoY2014, WangC2021, LiuH2021, WangJ2023}. 
As a result, symmetry requirements of nonlinear Hall effects are also different from those of linear Hall effects. 
Recent work on uniaxially strained graphene and organic semiconducting monolayer demonstrates that nonlinear valley Hall effect appears in centrosymmetric monolayers \cite{Das2024}. 
Compared with the above centrosymmetric monolayers, MX$_{2}$ bilayers possess an additional layer degree of freedom and nonnegligible spin-orbit coupling. 
Given these intriguing characteristics and vanishing linear valley Hall effects in MX$_{2}$ bilayers, it is worth studying their nonlinear valley Hall response to the longitudinal electric field.

In this work, taking the MoS$_{2}$ bilayer as an example, we study nonlinear valley Hall (NVH) effect in MX$_{2}$ bilayers, by combining low-energy effective models and first-principles calculations. 
Our model calculations demonstrate that a nonvanishing NVH conductivity is obtained by introducing the tilting into the Dirac cones via a uniaxial strain. 
The NVH conductivity originates from an intrinsic band property, i.e. Berry connection polarizability.
Compared with the valence band, a smaller spin-orbit-coupling-induced band splitting in the conduction band generates much remarkable NVH conductivity, which can be used as a detector for such small band splitting. 
Different from centrosymmetric monolayers, a van der Waals gap in the bilayer system is found to be an experimentally available knob for modulating the NVH conductivity. 
Besides, the strength and direction of the uniaxial strain also make the Hall conductivity highly tunable.  
The first-principles calculations of the uniaxially strained MoS$_{2}$ bilayer are further performed and corresponding results support the findings from effective models. 
Our work broadens material choices of valley materials with valley-contrasting transport properties and paves the way for designing advanced valleytronic devices based on nonlinear valley transports.

\begin{figure}[htb]
\includegraphics[width=8.0 cm]{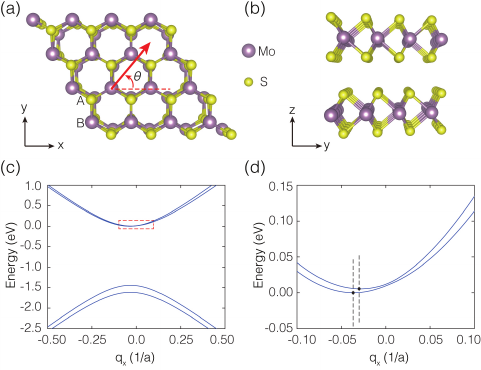}
\caption{ 
Crystal and band structures of a uniaxially strained 2H-MX$_2$ bilayer. 
(a) The top view and (b) the side view of the crystal structure. 
Purple and yellow balls stand for M and X atoms, respectively. 
The red arrow denotes the strain direction with a polar angle $\theta$. 
(c) Low-energy band structure at $K_{+}$ valley of a MoS$_2$ bilayer with a 5\% uniaxial strain along the $x$ direction, where the conduction band minimum is set to zero energy and $\bm{q}$ is given in units of the reciprocal of the lattice constant, $a$. 
(d) The magnified band structure around the conduction band edges at $K_{+}$ valley, corresponding to the red boxed region in (c). 
Black dots denote the minima of two conduction bands.  
The bands are found to be asymmetric with respect to perpendicular lines passing through the conduction band minima. 
}
\label{fig01}
\end{figure}

{\color{blue}\textit{Crystal structure and symmetry analyses.}} -- 
Figs. \ref{fig01} (a) and (b) show the top and side views of the MX$_{2}$ bilayer in the 2H phase, respectively. 
Since the 2H phase is abundant and the most thermodynamically stable for MX$_{2}$, the MX$_{2}$ bilayers in the 2H phase have been readily fabricated in experiments \cite{Strachan2021}. 
In the 2H phase of the MX$_{2}$ bilayer, upper and lower monolayers are superposed, but with a relative rotation of $\pi$ around the $z$-axis. 
That is, transition-metal (chalcogen) atoms of the upper monolayer are directly placed on chalcogen (transition-metal) atoms of the lower monolayer. 
The bilayer has a $D_{3d}$ point group symmetry, of which the generators include three-fold rotation, $C_{3z}$, along the $z$-axis, two-fold rotation, $C_{2x}$, along the $x$-axis, and the mirror, $M_{x}$, perpendicular to the $x$-axis. 
Besides, the spatial inversion symmetry $\mathcal{P}$ is also included in the point group, and the time-reversal symmetry $\mathcal{T}$ is preserved when a nonmagnetic MX$_{2}$ is considered here.

\begin{figure}[htb]
\includegraphics[width=8.5 cm]{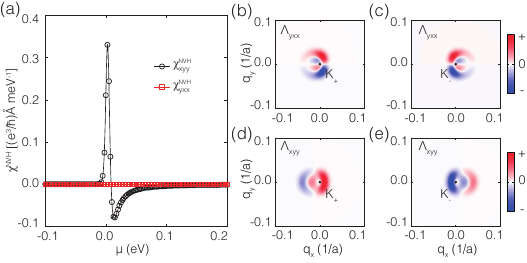}
\caption{ The NVH effect of a MoS$_2$ bilayer with a $2\%$ uniaxial strain along the $x$-axis. 
(a) The evolutions of $\chi^{\text{NVH}}_{xyy}$ and $\chi^{\text{NVH}}_{yxx}$ as functions of the chemical potential, where the energy window is chosen around the conduction band edge and the conduction band minimum is set to zero energy. 
(b)-(e) demonstrate the momentum-resolved $\Lambda_{yxx}$ near $K_{+}$, $\Lambda_{yxx}$ near $K_{-}$, $\Lambda_{xyy}$ near $K_{+}$, and $\Lambda_{xyy}$ near $K_{-}$, respectively. 
The black dots denote the locations of $K_{\pm}$ with coordinates, $q_{x}=q_{y}=0$. 
}
\label{fig02}
\end{figure}

According to above symmetries, the symmetry analyses of the NVH effects are further performed. 
The NVH effects can be described by the relation, 
\begin{equation}
J^{\text{NVH}}_{\alpha} = J^{K_{+}}_{\alpha} - J^{K_{-}}_{\alpha} = \chi^{\text{NVH}}_{\alpha\beta\gamma} E_{\beta}E_{\gamma}. 
\label{eq01}
\end{equation}
Here, $J^{\text{NVH}}_{\alpha}$ is NVH current, $\chi^{\text{NVH}}_{\alpha\beta\gamma}$ is corresponding conductivity tensor element, and $J^{K_{\pm}}_{\alpha}$ are nonlinear Hall contributions from $K_{\pm}$ valleys.  
$E_{\beta}$ and $E_{\gamma}$ are components of applied electric field. 
$\alpha$, $\beta$, and $\gamma$ are Cartesian coordinates. 
For a two-dimensional system, the coordinates are restricted to the $x$-$y$ plane. 
Applying the $\mathcal{T}$ or $\mathcal{P}$ symmetry operator on the quantities of Eq. \ref{eq01}, it is found that the equation still holds with nonzero $\chi^{\text{NVH}}_{\alpha\beta\gamma}$. 
It is because under the $\mathcal{P}$ or $\mathcal{T}$ operation, $J^{K_{\pm}}_{\alpha}$ is changed into $-J^{K_{\mp}}_{\alpha}$, leading to an invariant $J^{\text{NVH}}_{\alpha}$ on the left-hand side of the equation, and the product of two electric field components on the right-hand side is also invariant though $E_{\beta,\gamma}$ is reversed under the $\mathcal{P}$ operation. 
In contrast, the simultaneous presence of $\mathcal{P}$ and $\mathcal{T}$ symmetries excludes nonvanishing Berry curvatures and associated dipole \cite{Xiao2010, Sodemann2015}. 
Therefore, the MX$_{2}$ bilayer with $\mathcal{P}$ and $\mathcal{T}$ symmetries allows for the NVH effect arising from only the Berry connection polarizability (BCP), similar to centrosymmetric monolayers \cite{Das2024}. 
When applying the $C_{3z}$ operation, $\chi^{\text{NVH}}_{\alpha\beta\gamma}$ is vanishing to maintain the Eq. \ref{eq01}, since two sides of Eq. \ref{eq01} include one and two vectors, respectively, resulting in different phase changes after the rotation. 
The  $C_{3z}$ symmetry is thus not favored for generating nonlinear Hall currents, and correspondingly a uniaxial strain is applied to break the symmetry in the MX$_{2}$ bilayer, as shown by the red arrow in Fig. \ref{fig01} (a), where the polar angle $\theta$ denotes the strain direction with respect to the zigzag direction ($x$-axis). 
Further considering the setup of two-dimensional Hall transport experiment, there are two possible conductivity tensor elements, $\chi^{\text{NVH}}_{xyy}$ and $\chi^{\text{NVH}}_{yxx}$. 
Under the strain along the $x$ or $y$ direction, the presences of both $C_{2x}$ and $M_{x}$ symmetries allow for nonzero $\chi^{\text{NVH}}_{xyy}$ but exclude $\chi^{\text{NVH}}_{yxx}$, which will become clear later by momentum-resolved BCP in Fig. \ref{fig02}.

{\color{blue}\textit{Model calculations of a MX$_2$ bilayer.}} -- 
Armed with the symmetry analyses, we performed effective model calculations of uniaxially strained MX$_{2}$ bilayers. 
Using $\{\psi^{\tau}_{lc}, \psi^{\tau}_{lv}, \psi^{\tau}_{uc}, \psi^{\tau}_{uv}\}\otimes\{\uparrow, \downarrow\}$ as a basis set, a spinful four-orbital $\bm{k}\cdot \bm{p}$ Hamiltonian, $\mathcal{H}^{\tau}$, in the neighborhood of $K_{\pm}$ valleys reads,  
 \begin{equation}
\mathcal{H^{\tau}} = 
\begin{pmatrix}
H^{\tau}_{l} & H^{\tau}_{lu} \\ 
H^{\tau}_{ul} & H^{\tau}_{u} 
\end{pmatrix}. 
\label{eq02}
 \end{equation} 
 In the basis set, the superscript $\tau=\pm 1$ denotes $K_{\pm}$ valleys. 
 The subscripts, $l$ and $u$, represent lower and upper MX$_{2}$ monolayers, respectively, while $c$ and $v$ stand for the conduction and valence bands. 
 $\psi^{\tau}_{lc}$ is thus the wave function at $\tau$ valley, contributed by the conduction band of the lower monolayer, and the other wave functions can also be well understood in the same manner.  
 $\uparrow$ and $\downarrow$ denote opposite spins. 
 In the Hamiltonian, $H_{u}^{\tau}$ and $H_{l}^{\tau}$ are the effective models of the upper and lower monolayers, respectively, while $H_{lu}^{\tau}$ and $H_{ul}^{\tau}$ represent interlayer interactions. 
 
 With the layer index $\eta=\pm 1$ distinguishing different monolayers, the model of the single monolayer is written as, \cite{Liu2013, Fang2018}
 \begin{eqnarray} 
 &&H^{\tau}_{l/u} = \hbar v_{F} (\tau q_{x} \sigma_{x} - \eta q_{y} \sigma_{y}) s_{0} +  \frac{\Delta}{2}\sigma_{z} s_{0} + \delta\sigma_{0}s_{0}  \nn  \\
 &&  + \lambda_{c} \tau \eta \frac{(\sigma_{0}+\sigma_{z})}{2} s_{z} + \lambda_{v} \tau \eta \frac{(\sigma_{0}-\sigma_{z})}{2} s_{z}   \nn \\
 &&  + \epsilon [\gamma_{1}\sigma_{0} + \gamma_{2} \sigma_{z}+\gamma_{3} (\cos 2\theta\sigma_{x} - \tau\eta \sin 2 \theta  \sigma_{y})]  s_{0}. 
 \label{eq03}
 \end{eqnarray}
The three lines of $H_{l/u}^{\tau}$ describe gapped Dirac fermion, the spin-orbit interaction, and the uniaxial strain along the $\theta$ direction, respectively. 
Herein, $q_{x}$ and $q_{y}$ are in-plane crystal wave vectors with respect to $K_{\tau}$. 
$\sigma_{x,y,z}$ and $s_{z}$ are Pauli matrices for orbital and spin spaces, respectively, while $\sigma_{0}$ and $s_{0}$ are corresponding identity matrices. 
$\hbar$, $v_{F}$, $\Delta$, and $\delta$ are the reduced Planck constant, Fermi velocity, band splitting from the crystal field, and immaterial energy shift, respectively. 
$\lambda_{c}$ and $\lambda_{v}$ are the spin-orbit coupling (SOC) strengths of two bands. 
With $\epsilon$ and $\gamma_{1-3}$ being the applied strain and strain-related parameters, respectively, the strain terms are obtained by assuming a zero Possion ratio. 
On the other hand, the interlayer interaction reads \cite{Kormanyos2018}
 \begin{equation}
 H^{\tau}_{lu} = (H^{\tau}_{ul})^{\dagger} = 
 \begin{pmatrix}
t_{cc}(\tau q_{x} + i q_{y}) & 0 \\ 
0 & t_{vv} 
\end{pmatrix}, 
\end{equation} 
where $t_{cc}$ $(t_{vv})$ is an interlayer hopping strength between conduction (valence) band states from two monolayers. 

By adopting parameters of MoS$_{2}$ \cite{Liu2013, Fang2018, Kormanyos2018}, which are provided in Supplementary Material (SM hereafter), the band structure of the uniaxially strained MoS$_{2}$ bilayer is calculated based on the Hamiltonian \ref{eq02} and shown in Fig. \ref{fig01} (c). 
It is seen that there are two conduction bands and two valence bands, which are all doubly degenerate due to the simultaneous presence of $\mathcal{P}$ and $\mathcal{T}$ symmetries. 
Band splittings appear between two conduction (valence) bands. 
The splittings between valence band edges arise from the combination roles of the SOC denoted by $\lambda_{v}$ and the interlayer hopping $t_{vv}$, while those between conduction band edges are mainly determined by the SOC strength $\lambda_{c}$. 
Since $\lambda_{v}$ and $t_{vv}$ are much larger than $\lambda_{c}$, the band splittings between valence bands, with magnitudes on the order of 100 meV, are larger than those between conduction bands by the order of magnitude.   
Under the applied strain, the edges of all bands are found to move away from the $K_{\pm}$ points with coordinates, $q_{x}=q_{y}=0$, and the rotation symmetry is broken at valleys, compared with an isotropic band structure without strain. 
Focusing on the conduction bands within a small energy window in Fig. \ref{fig01} (d), it is seen that these bands are also no longer symmetric with respect to perpendicular lines that pass through the conduction band minima, indicating the appearance of band tilts.

Considering that the tilted band is likely to generate nonlinear Hall currents \cite{Du2018, WangC2021, LiuH2021, WangJ2023, Das2024}, we then calculate the NVH conductivity of the MoS$_{2}$ bilayer. 
The conductivity tensor element is defined as \cite{Das2024}
\begin{eqnarray}
\chi^{\text{NVH}}_{\alpha\beta\gamma} &=& \frac{e^{3}}{\hbar} \int_{\text{BZ}} \frac{d\bm{k}}{(2\pi)^{d}} (-1)^{\tau}  \Lambda_{\alpha\beta\gamma}(\bm{k})  \nn \\
&=& \frac{e^{3}}{\hbar} \int_{\text{BZ}} \frac{d\bm{k}}{(2\pi)^{d}} (-1)^{\tau} \sum_{n} \lambda_{\alpha\beta\gamma}^{n}(\bm{k}) \frac{\partial f(\varepsilon_{n\bm{k}})}{\partial \varepsilon_{n\bm{k}}}, 
\label{eq05}
\end{eqnarray}
with \begin{eqnarray}
\lambda^{n}_{\alpha\beta\gamma}(\bm{k}) &=&  v^{n}_{\alpha}G^{n}_{\beta\gamma}(\bm{k})-v^{n}_{\beta}G^{n}_{\alpha\gamma}(\bm{k}), \\  \nn \\
G_{\alpha\beta}^{n}(\bm{k}) &=& 2 \text{Re} \sum_{m\ne n} \frac{A^{nm}_{\alpha}(\bm{k})A^{mn}_{\beta}(\bm{k})}{\varepsilon_{n\bm{k}}-\varepsilon_{m\bm{k}}}. 
\label{eq07}
\end{eqnarray} 
Here, $G_{\alpha\beta}^{n}$ is the BCP. 
$\Lambda_{\alpha\beta\gamma}$  is the BCP dipole, while $\lambda_{\alpha\beta\gamma}^{n}$ is the band-resolved one.  
 $A^{nm}_{\alpha} (\bm{k})=\langle u_{n\bm{k}} | i\partial_{\alpha} | u_{m\bm{k}} \rangle$ is the interband Berry connection, with $u_{n\bm{k}}$ the periodic part of the Bloch wave function of the $n$th band at the wave vector $\bm{k}$. 
$\varepsilon_{n\bm{k}}$ is the corresponding energy of the Bloch state. 
 $f(\varepsilon_{n\bm{k}})$ is the equilibrium Fermi-Dirac distribution function. 
  $v_{\alpha}^{n}$ is the velocity expectation. 
$(-1)^{\tau}$ is introduced into Eq. \ref{eq05} to distinguish contributions from different valleys.

Fig. \ref{fig02} (a) shows calculated conductivity tensor elements, $\chi_{xyy}$ and $\chi_{yxx}$, of the MoS$_{2}$ bilayer with a $2\%$ uniaxial strain along the $x$ axis, as functions of the chemical potential, $\mu$. 
It is seen that while $\chi_{yxx}$ is always zero,  $\chi_{xyy}$ is nonvanishing. 
Compared with the case of the valence bands, the magnitude of $\chi_{xyy}$ at the conduction band edge is much larger. 
Therefore, the chemical potential is chosen around the conduction band edge in Fig. \ref{fig02} (a), while the result of the valence bands is provided in SM.   
When lifting the chemical potential from the conduction band edge, $\chi_{xyy}$ first increases sharply and forms a high positive peak, and then reverses the sign along with a low negative peak, and finally decreases to zero. 
The calculated $\chi_{xyy}$ has a maximum value of 0.33 $(e^{3}/\hbar)$\AA\ meV$^{-1}$, i.e. $\sim 8.03$ nm$\cdot$ mA/V$^{2}$ at $\mu=3$ meV, which is much larger than the NVH conductivity of the strained graphene, comparable to the NVH conductivity of the organic conductor $\alpha$(BEDT-TTF)$_{2}$I$_{3}$ \cite{Das2024} and also surpass the nonlinear Hall conductivity of few-layer WTe$_{2}$ that has been observed experimentally \cite{Kang2019, WangH2019}. 
Therefore, a considerable NVH conductivity is expected to be detected in the strained MoS$_{2}$ bilayer.

To elucidate the disparity between $\chi_{xyy}$ and $\chi_{yxx}$, the momentum-resolved BCP dipoles, $
\Lambda(\bm{k})$, at $K_{\pm}$ valleys are provided in Fig. \ref{fig02} (b)-(e). 
For $\Lambda_{yxx}$ at a certain valley, it is mirror-antisymmetric with respect to the line $k_{y}=0$, leading to vanishing nonlinear Hall conductivity from each valley. 
Accordingly, the NVH conductivity is zero as well.  
In contrast, $\Lambda_{xyy}$ is mirror-symmetric with respect to $k_{y}=0$, allowing for nonvanishing nonlinear Hall conductivity from each valley. 
Comparing $\Lambda_{xyy}$ at two valleys, opposite $\Lambda_{xyy}$ enable the emergence of NVH current.  
The patterns of $\Lambda(\bm{k})$ at $K_{\pm}$ valleys can be well understood by the crystal symmetry. 
As mentioned above, two-fold rotation $C_{2x}$ is a symmetry of the bilayer. 
The symmetry can be regarded as a mirror line $k_{y}=0$ at each valley in two-dimensional momentum space.  
On the other hand, $\Lambda_{\alpha\beta\beta}(\bm{k})$ at $K_{\tau}$ valley is proportional to $J_{\alpha}^{K_{\tau}}(\bm{k})/ E^{2}_{\beta}$ according to Eq. \ref{eq01}. 
  When $E^{2}_{\beta}$ is invariant under a symmetry operation, e.g. $C_{2x}$ and $M_{x}$, the transformation of $\Lambda_{\alpha\beta\beta}(\bm{k})$ at $K_{\tau}$ valley is determined by $J_{\alpha}^{K_{\tau}}(\bm{k})$. 
Since $C_{2x}J^{K_{\tau}}_{x}(q_{x},q_{y})=J^{K_{\tau}}_{x}(q_{x},-q_{y})$ and $C_{2x}J^{K_{\tau}}_{y}(q_{x},q_{y})=-J^{K_{\tau}}_{y}(q_{x},-q_{y})$, $\Lambda_{xyy}$ and $\Lambda_{yxx}$ are mirror-symmetric and mirror-antisymmetric at each valley, respectively. 
Further considering that two valleys can be related by the $M_{x}$ symmetry, the relation, $M_{x} J^{K_{\tau}}_{x}(q_{x},q_{y})=-J^{K_{-\tau}}_{x}(-q_{x},q_{y})$, results in opposite nonlinear Hall currents from two valleys and nonvanishing NVH conductivity. 
In addition to the above analyses, it is also noted that the mirror line $k_{y}=0$ at each valley can also be derived from the $M_{x}\mathcal{T}$ symmetry like the $C_{2x}$ symmetry, and the connection between two valleys can also be realized by the $\mathcal{T}$ or $\mathcal{P}$ symmetry, which will reach the same conclusion.

Besides the above BCP dipole, $\Lambda_{xyy}(\bm{k})$, band-resolved contribution, $\lambda_{xyy}^{n}(\bm{k})$, and $\partial f(\varepsilon_{n\bm{k}})/ \partial \varepsilon_{n\bm{k}}$ manifesting the Fermi surface are also provided in SM. 
Compared with the case of the pristine bilayer, the introduction of the uniaxial strain makes $\partial f(\varepsilon_{n\bm{k}})/ \partial \varepsilon_{n\bm{k}}$ and associated Fermi surface non-circular and off-center, indicating the anisotropy of the tilted bands. 
There is also an increased disparity between magnitudes of positive and negative contributions from different $\bm{k}$ to $\lambda_{xyy}^{n}(\bm{k})$. 
The increased disparity of $\lambda_{xyy}^{n}(\bm{k})$ and anisotropic $\partial f(\varepsilon_{n\bm{k}})/ \partial \varepsilon_{n\bm{k}}$ lead to sizable $\Lambda_{xyy}(\bm{k})$ and corresponding NVH conductivity.

\begin{figure}[htb]
\includegraphics[width=8.5 cm]{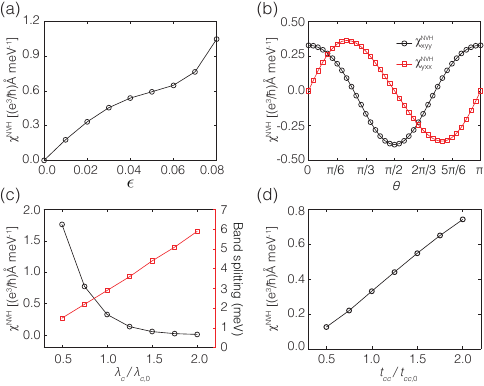}
\caption{ 
Evolutions of first peak values of NVH conductivities. 
(a) The evolution of $\chi_{xyy}^{\text{NVH}}$ as a function of the strain, where $\theta=0$ is adopted. 
(b) The evolutions of $\chi_{xyy}^{\text{NVH}}$ and $\chi_{yxx}^{\text{NVH}}$ as functions of the strain direction, where a 2\% strain is adopted. 
(c) The evolution of $\chi_{xyy}^{\text{NVH}}$ and band splitting minimum as functions of the SOC strength, $\lambda_{c}$, in units of the value of the MoS$_{2}$ bilayer, $\lambda_{c,0}$=$-1.5$ meV. 
(d) The evolution of $\chi_{xyy}^{\text{NVH}}$ as a function of the interlayer hopping strength, $t_{cc}$, in units of the value of the MoS$_{2}$ bilayer, $t_{cc,0}$=$70.6$ meV$\cdot$\AA. 
In (c)-(d), a 2\% strain with $\theta=0$ is applied.  
}
\label{fig03}
\end{figure}

{\color{blue}\textit{Highly tunability of NVH conductivities}} --Given that the nonvanishing NVH conductivity results from the strain-induced band tilts, the magnitude of the strain is first expected to modulate NVH effect. 
As illustrated in Fig. \ref{fig03} (a), as the applied strain increases, $\chi_{xyy}^{\text{NVH}}$ is continuously enhanced.  
It can also be well explained by the more anisotropic $\partial f(\varepsilon_{n\bm{k}})/ \partial \varepsilon_{n\bm{k}}$  and continuously increased disparity between positive and negative  $\lambda_{xyy}^{n}(\bm{k})$ (see SM). 
Besides the strength of applied strain, the strain direction is likely to be another experimental accessible knobs to tune the NVH conductivity. 
As mentioned above, when the uniaxial strain is along $x$ and $y$ direction, the presences of $C_{2x}$ and $M_{x}\mathcal{T}$ symmetries exclude nonzero $\chi_{yxx}^{\text{NVH}}$. 
By moving the strain direction away from $x$ and $y$ axes, the symmetry constraints are removed, and $\chi_{yxx}^{\text{NVH}}$ becomes possible. 
Fig. \ref{fig03} (b) shows the evolutions of $\chi_{xyy}^{\text{NVH}}$ and $\chi_{yxx}^{\text{NVH}}$ as functions of the strain direction, $\theta$. 
It is seen that $\chi_{yxx}^{\text{NVH}}$ is indeed nonvanishing with varied $\theta$. 
$\chi_{xyy}^{\text{NVH}}$ and $\chi_{yxx}^{\text{NVH}}$ behave like cosine and sine functions of $2\theta$, respectively. 
Therefore, the strength of the strain can be employed to modulate the magnitude of NVH conductivity, and the strain direction is capable of regulating both magnitudes and signs of NVH conductivities.

Furthermore, going back to intrinsic properties of the MoS$_2$ bilayer, e.g. the band splittings and layer degree of freedom, we explore more approaches for tuning the NVH effect. 
As mentioned above, the conduction bands possess much larger $\chi_{xyy}^{\text{NVH}}$, compared with the valence bands. 
The NVH conductivity can thus be regulated by the shift of the chemical potential via carrier doping.  
The large $\chi_{xyy}^{\text{NVH}}$ of the conduction bands can be interpreted by the definition of the BCP in Eq. \ref{eq07}. 
Since $A_{\alpha}^{nm} \propto (\varepsilon_{n\bm{k}}-\varepsilon_{m\bm{k}})^{-1}$ and accordingly $G_{\alpha\beta}^{n} \propto (\varepsilon_{n\bm{k}}-\varepsilon_{m\bm{k}})^{-3}$, the NVH conductivity is highly sensitive to magnitudes of band splittings. 
The smaller band splitting between the conduction bands helps to enhance the NVH effect, compared with the larger splittings between the valence bands. 
 To confirm the connection between the band splitting and NVH conductivity, the SOC strength, $\lambda_{c}$, of the conduction bands is changed to tune the band splittings between conduction bands in Fig. \ref{fig03} (c). 
 The minimum of the band splittings is indeed found to reduce as $\lambda_{c}$ decreases, leading to an increasing NVH effect.   
 Therefore, in order to enhance the NVH conductivity, the chemical potential is expected to be regulated to the energy where a small band gap is located.  
 On the other hand, the considerable NVH effect can serve as a detectable signal for such small band splittings between conduction bands.

Moreover, as a van der Waals bilayer system, the interlayer gap, which is inversely proportional to the interlayer interaction, is a unique and experimentally adjustable characteristic, distinct from monolayers \cite{Chen2024}.  
 The effect of the interlayer hopping strength, $t_{cc}$, is thus investigated and shown in Fig. \ref{fig03} (d).  
 $\chi_{xyy}^{\text{NVH}}$ is enhanced with the increase of $t_{cc}$, indicating that NVH conductivity can be enlarged by narrowing the van der Waals gap. 
 The enhanced $\chi_{xyy}^{\text{NVH}}$ can also be related to the more anisotropic $\partial f(\varepsilon_{n\bm{k}})/ \partial \varepsilon_{n\bm{k}}$ and increased disparity of $\lambda_{xyy}^{n}(\bm{k})$, as provided in SM. 
 
 In addition to the above modulation methods, a two-orbital effective model and an analytic expression of $\chi_{xyy}^{\text{NVH}}$ are further derived for conduction bands to further reveal the origin of these modulation methods. 
 The derivation and related discussions are given in SM.

\begin{figure}[htb]
\includegraphics[width=8.5 cm]{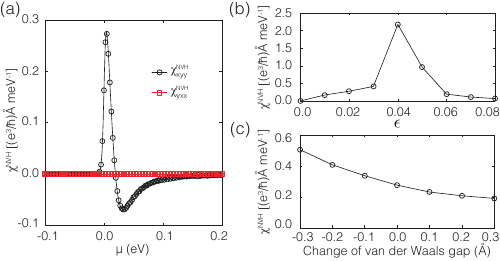}
\caption{ First-principles NVH conductivities of a MoS$_2$ bilayer with a uniaxial strain along the $x$-axis. 
(a) The evolutions of $\chi^{\text{NVH}}_{xyy}$ and $\chi^{\text{NVH}}_{yxx}$ as functions of the chemical potential, where the energy window is chosen around the conduction band edge and the conduction band minimum is set to zero energy. 
(b) The evolution of the first peak value of $\chi_{xyy}^{\text{NVH}}$ as a function of the strain. 
(c) The evolution of the first peak value of $\chi_{xyy}^{\text{NVH}}$ as a function of the van der Waals gap with respect to the fully relaxed one, $d_{0}=3.09$ \AA. 
In (a) and (c), a $2\%$ strain is applied. 
}
\label{fig04}
\end{figure}

{\color{blue}\textit{First-principles calculations of NVH conductivities}} --
To verify the findings based on effective models, first-principles calculations of NVH conductivities are further performed for the MoS$_{2}$ bilayer. 
The details of first-principles calculations can be found in SM. 
Fig. \ref{fig04} (a) presents the evolutions of $\chi_{xyy}^{\text{NVH}}$ and $\chi_{yxx}^{\text{NVH}}$ with varied chemical potential, where the applied strain is the same as that in the model calculation of Fig. \ref{fig02} (a).
It is seen that there is considerable $\chi_{xyy}^{\text{NVH}}$ and vanishing $\chi_{yxx}^{\text{NVH}}$.   
 The change trend of $\chi_{xyy}^{\text{NVH}}$ is very similar to that in Fig. \ref{fig02} (a), and the positive extreme of $\chi_{xyy}^{\text{NVH}}$ has a value of $0.28$ $(e^{3}/\hbar)$\AA\space meV$^{-1}$ that is close to the model calculation as well. 
 Fig. \ref{fig04} (b) presents the evolution of the positive extreme of $\chi_{xyy}^{\text{NVH}}$ with the strain $\epsilon$. 
 $\chi_{xyy}^{\text{NVH}}$ first increases under a small strain, $\epsilon\le 4\%$, agreeing with the model calculation in Fig. \ref{fig03} (a). 
 Then $\chi_{xyy}^{\text{NVH}}$ decreases when $\epsilon> 4\%$. 
 The nonmonotonic variation is also embodied in analytic expression of $\chi_{xyy}^{\text{NVH}}$ in SM. 
 In contrast, the nonmonotonicity is absent in the model calculation for the strain range considered in Fig. \ref{fig03} (a), which may result from ignoring $\bm{q}$-related high-order terms in the Hamiltonian \ref{eq02}.  
 Moreover, with the fully relaxed van der Waals gap as a reference, the gap is changed by -0.3 \AA\space to +0.3 \AA\space in Fig. \ref{fig04} (c). 
 As the van der Waals gap increases, $\chi_{xyy}^{\text{NVH}}$ is found to continuously decrease. 
 Therefore, the NVH conductivity can be indeed modulated via the van der Waals gap in a bilayer system, with the help of the layer degree of freedom.

{\color{blue}\textit{Summary}} --
To conclude, a NVH effect of bilayer MoS$_{2}$ has been investigated by effective models and first-principles calculations. 
The NVH conductivity is considerable in the vicinity of small spin-orbit splittings between conduction bands, and it is highly tunable by applied strain, chemical potential, and intrinsic layer degree of freedom. 
The nonlinear valley transport is expected to be detected experimentally by nonlocal resistance measurement \cite{Sui2015, Shimazaki2015}.  
Moreover, given that MoS$_{2}$ is one example of a large family of transition metal dichalcogenides \cite{Manzeli2017, Choi2017, Wu2021},  the NVH conductivity found in MoS$_{2}$ bilayer is likely to have more generalizations to other MX$_{2}$ bilayers. 
Further considering a diversity of exotic physical properties in MX$_{2}$, our work provides a fertile platform for designer valleytronic devices based on nonlinear transport.

{\color{blue}\textit{Acknowledgments}} -- 
We appreciate helpful discussions with Huiying Liu. 
We are supported by the National Natural Science Foundation of China (Nos. 12374044, 12004186, 11904173). 

\appendix

%

\end{document}